\documentclass[aps,prl,twocolumn,superscriptaddress]{revtex4}
\usepackage{graphicx,epsf}
\usepackage{graphicx}

\begin{document}

\title{Alloying mechanisms for epitaxial nanocrystals}

\author{M. S. Leite}
\affiliation{Laborat\'orio Nacional de Luz S\'{\i}ncrotron, Caixa
Postal 6192 - CEP 13084-971, Campinas, SP, Brazil}
\affiliation{Instituto de F\'isica 'Gleb Wataghin', Universidade
Estadual de Campinas, Campinas, SP, Brazil}

\author{G. Medeiros-Ribeiro}
\email[corresponding author:]{gmedeiros@lnls.br}
\affiliation{Laborat\'orio Nacional de Luz S\'{\i}ncrotron, Caixa
Postal 6192 - CEP 13084-971, Campinas, SP, Brazil}
\affiliation{Hewlett-Packard Laboratories, 1501 Page Mill Rd., 94304
Palo Alto, CA}
\author{T. I. Kamins}
\affiliation{Hewlett-Packard Laboratories, 1501 Page Mill Rd., 94304
Palo Alto, CA}

\author{R. Stanley Williams}
\affiliation{Hewlett-Packard Laboratories, 1501 Page Mill Rd., 94304
Palo Alto, CA}

\date{\today}

\begin{abstract}

The different mechanisms involved in the alloying of epitaxial
nanocrystals are reported in this letter. Intermixing during growth,
surface diffusion and intra-island diffusion were investigated by
varying the growth conditions and annealing environments during
chemical vapor deposition. The relative importance of each mechanism
was evaluated in determining a particular composition profile for
dome-shaped Ge:Si (001) islands. For samples grown at a faster rate,
intermixing during growth was reduced. Si surface diffusion
dominates during H$_2$ annealing whereas Ge surface diffusion and
intra-island diffusion prevail during annealing in a PH$_3$
environment.

\end{abstract}

\maketitle

In coherently-strained epitaxial islands, the most important factor
that determines island size and stability is composition.
Composition variations inside epitaxial islands will substantially
influence their structural properties and, as a consequence, the
electronic properties of epitaxial nanocrystals\cite{Tersoff98}. The
composition profile of SiGe islands has only recently been
measured\cite{Malachias03,Denker2003,Floyed2003}, and its origin
depends on kinetic and thermodynamic contributions, which sometimes
are difficult to separate. In a careful and detailed
study\cite{Katsaros05}, some of the pathways for Si and Ge
intermixing have been investigated in Molecular Beam Epitaxy (MBE)
grown islands. A series of experiments was carried out at different
temperatures and with subsequent annealing steps in Ultra High
Vacuum (UHV). The main result for the reported experimental
conditions was that surface diffusion of Si and Ge was the dominant
mechanism determining the island composition profile. It is known,
however, that annealing and growth in different ambient conditions (i.e., UHV,
as opposed to H$_{2}$ or PH$_{3}$ environments) selectively changes
the surface mobility of adatoms\cite{Kamins03,Kamins04}.  As a
result, alloying in the presence of gases can proceed privileging
selected mechanisms during Chemical Vapor Deposition (CVD) growth.

Bulk diffusion requires the formation of vacancies and/or
interstitials\cite{fahey89}. For buried 2D SiGe layers, diffusion of
Ge was found to increase with the Ge content and compressive stress
\cite{zangenberg01}. Although the amount of diffusion inferred from
these results extrapolated to 600 $^\circ$C is negligible for
\emph{unstrained} material, the activation energies depend strongly on
strain. For the case of sub monolayer coverages, even at
temperatures around 500 $^{\circ}$C, Ge diffusion and intermixing
into Si surfaces has been theoretically predicted and was observed
by high resolution Rutherford Back Scattering (RBS) before the first
monolayer of material was completely
deposited\cite{Uberuaga00,Nakajima99}. In self-assembled islands, a
significant amount of strain is present and the fact that high index
facets and edges (and hence defects) make up the surface provides a
larger number of pathways for intermixing compared to a 2D film.
Alloying in coherently-strained nanocrystals needs to be
investigated in more detail to understand island formation and
evolution. In order to comprehend the composition profiles, it is
imperative to vary the kinetic and thermodynamic components
individually in experiments to elucidate the mechanisms that lead to
intermixing for a particular growth condition.

The primary mechanisms that modulate the composition profile of
self-assembled islands are: \textbf{a}) exchange reactions between
Si and Ge during island growth, defined by the attachment and
detachment of atoms between the crystal and 2D adatom gas;
\textbf{b}) surface diffusion of both Si and Ge adatoms; and
\textbf{c}) diffusion of Ge and Si atoms within the islands (intra-island diffusion),
excluding the surface adatoms. Inter-island
diffusion, which is a special case of surface diffusion, is not
covered here in detail, but is also a crucial component for the
final island composition profile. Changing the growth and annealing
conditions allows us to privilege one mechanism at a time, but not
rigourously supress the other two. The goal of this work is to
evaluate the relative importance of each mechanism by comparing
samples grown by CVD and annealed under different conditions. For instance,
comparing samples grown at different rates affects all mechanisms,
but more effectively \textbf{a} and \textbf{b}. Annealing in a
H$_{2}$ environment decreases surface mobility of both Ge and Si
compared to annealing in UHV\cite{Kamins03}, yet does not completely
stop the surface diffusion of either species. Thus it allows both
mechanisms \textbf{b} and \textbf{c} to be investigated. Annealing
in a PH$_{3}$ environment substantially reduces Si surface
diffusion, yet has little effect on Ge surface diffusion. This
different behavior occurs because the P-Si bond is stable (bond
enthalpy equal to 364 $\pm$ 7 kJ/mol) whereas the P-Ge bond is
unstable\cite{Handbook}. In this case, mechanism \textbf{c}
dominates for both Ge and Si species, and mechanism \textbf{b}
persists for Ge adatoms.

Four samples with nominally the same Ge deposition thickness of 12
eq-ML were grown at 600 $^{\circ}$C in a H$_{2}$ ambient in a
commercial CVD reactor on 150 mm diameter Si (001) wafers. The
conditions were chosen to produce dome shaped-islands. The
reproducibility of film thickness from run to run as determined by
RBS analysis was better than 5\%. The
first two samples were grown at $\it{P}$(GeH$_{4}$)=5x10$^{-4}$Torr
(as-deposited F-Fast - 6eq. ML/min) and
$\it{P}$(GeH$_{4}$)=2.5x10$^{-4}$Torr (as-deposited S-Slow - 3eq.
ML/min) in a 10 Torr ambient composed mainly of H$_{2}$ and
immediately cooled to room temperature. The other two samples were
grown at the same rate as sample S (reference sample); however,
after deposition of the Ge film they were subsequently annealed
\textit{in-situ} for 10 minutes at the growth temperature
(600$^\circ$C) in PH$_{3}$/H$_{2}$ with up to 1.4x10$^{-5}$ Torr
added PH$_{3}$ (annealed P) or H$_{2}$ (annealed H) environments.
The samples were characterized initially by Reciprocal-Space Mapping
(RSM) using a conventional CuK$\alpha$ X-ray tube using the (224)
and (004) reflections in order to extract the average Ge
content\cite{Sluis}. Selective chemical etching in
25\%NH$_{4}$OH:31\%H$_{2}$O$_{2}$ room temperature solution for
varying times was used to study the composition profiles in more
detail. This etchant is known to be more sensitive to the Ge
concentration variation than the RSM measurements, and to slowly
remove Ge-rich SiGe alloys with exponentially varying Ge selectivity
\cite{Denker05,Katsaros06}, allowing a detailed study of the
remaining material. Although the absolute composition obtained from
this technique is not known with a high precision as Anomalous X-ray
diffraction \cite{Malachias03,Schulli03} or Electron-energy-loss
spectroscopy (EELS) in a scanning transmission electron microscope
(STEM)\cite{Mcdaniel05,Vanfleet07}, the relative comparisons between
samples are far more sensitive than either method. For this work our
conclusions rely primarily on the relative comparisons rather than
on the knowledge of the absolute content. Local and statistical
analysis were performed for the as-grown and etched samples with
Atomic Force Microscopy (AFM) over ensembles of about 400 islands
per etching condition.

Table 1 shows a summary of the growth parameters used and the
average Ge composition obtained through the RSM experiments. Samples
as-deposited F, as-deposited S and annealed P were found to have the
same average Ge content within the experimental uncertainties.
However, a lower Ge content was found for sample annealed H,
accompanied by a broader diffraction peak corresponding to a wider
distribution of compositions within the island ensemble (not shown).
These results suggest that Si surface diffusion plays an important
role in the final composition profile for sample H, whereas for the
other samples that mechanism is minimal. This also confirms the low
surface diffusivity of Si in a PH$_{3}$ environment. The last column
displays the total integrated amount of material in the islands,
which is consistent with the total amount of Ge deposited of
12.0 $\pm$ 0.5ML and the 3.5 ML thick wetting layer.

\begin{table}[ht]
\caption{Growth parameters and average Ge content in each sample
obtained through reciprocal-space mapping using a conventional X-ray
tube. The integrated thickness corresponds to the total island
volume material integrated per area, not including the wetting
layer. The samples were grown by CVD, at 600 $^{\circ}$C. }

\begin{tabular}{l|p{1.8cm}|c|c|p{1.8cm}}
\hline Sample & growth rate (ML/min) & Annealing & $<$Ge\%$>$ & Integrated thickness (ML)\\

\hline
as-deposited F & 6 & No & 64 $\pm$ 5 & 6.5 $\pm$ 1.5\\
as-deposited S & 3 & No & 63 $\pm$ 5& 7.0 $\pm$ 1.5\\
annealed P & 3 & 10' PH$_3$ & 64 $\pm$ 5 & 11 $\pm$ 1.5\\
annealed H & 3 & 10' H$_2$ & 53 $\pm$ 5 & 6.5 $\pm$ 1.5\\

\hline
\end{tabular}
\label{emap}
\end{table}

Figure 1 shows the evolution of samples as-deposited F and S,
annealed P and H, before and after etching for 30 min and 60 min.
The top images correspond to 250nm $\times$  250nm AFM scans, and
the bottom graphs correspond to line profiles taken on statistically
representative islands selected from height histograms. The different
profiles indicate different degrees of alloying. After 4 hours of
etching the entire island material was removed for all cases,
leaving a visible moat around the region where the islands were
previously located (not shown).

Comparing the AFM images and the line scans before etching, we found
that the domes of sample annealed H are slightly larger in diameter;
their height is the same as for the dome islands of sample
as-deposited S. By comparing samples as-deposited S and annealed P,
three main observations can be made for sample P: a) the domes are
slightly taller, b) the islands total integrated volume is
significantly larger and c) there are no pyramids. From these
observations we can conclude that in addition to Ge inter-island
diffusion, material from the substrate is effectively being
incorporated into the islands causing their growth (comparing samples S and P this amounts to roughly 4 $\pm$ 3 ML). This has been observed recently also by STEM-EELS experiments in samples annealed
at 650$^\circ$C\cite{Vanfleet07}, producing a non-abrupt yet uniform
interface.

All but sample annealed H consisted of a symmetric Ge-rich outer
shell as shown previously by Grazing Incidence Anomalous X-Ray
Diffraction (GIXD) experiments\cite{Malachias03}. For sample H an irregular
composition profile is revealed by the selective etching, as seen in Figure 2.
This has been observed previously in CVD grown samples\cite{Kamins99},
and more recently for MBE grown material\cite{Denker05}, and both are consistent
with a a significant amount of Si surface diffusion and concomitant intermixing.
This asymmetric alloying profile is in accord with the GIXD results of  wider composition range within the
islands. For sample H, Si and Ge surface diffusion (mechanism
\textbf{b}) and possibly intra-island diffusion (mechanism
\textbf{c}) act together producing the observed morphology.

For all etching times, sample as-deposited F exhibited a larger
fraction of removed material, indicating that intermixing during
growth (mechanism \textbf{a}) as well as the other mechanisms are
reduced by the shorter growth time. A recent report
\cite{Mcdaniel05} on MBE grown material portrayed similar results.

After 30 minutes of etching, the domes in sample as-deposited S have
a top that is richer in Ge compared to sample annealed P. After 60
minutes of etching, one finds that more material has been removed
from sample P in contrast to sample as-deposited S and annealed H,
indicating a Ge rich region in sample P. These facts demonstrate
that Ge and Si redistribute inside the island during annealing in
PH$_{3}$, with Ge diffusing down towards the island base, and Si
diffusing up towards the island apex. It is interesting to compare
the inferred profile to results of Monte Carlo simulations, which
allow for intra-island diffusion\cite{Hadjisavvas05,lang05}; both
experiment and simulation show a SiGe core and a Ge rich shell.

\begin{figure}[ht]
\centerline{\epsffile{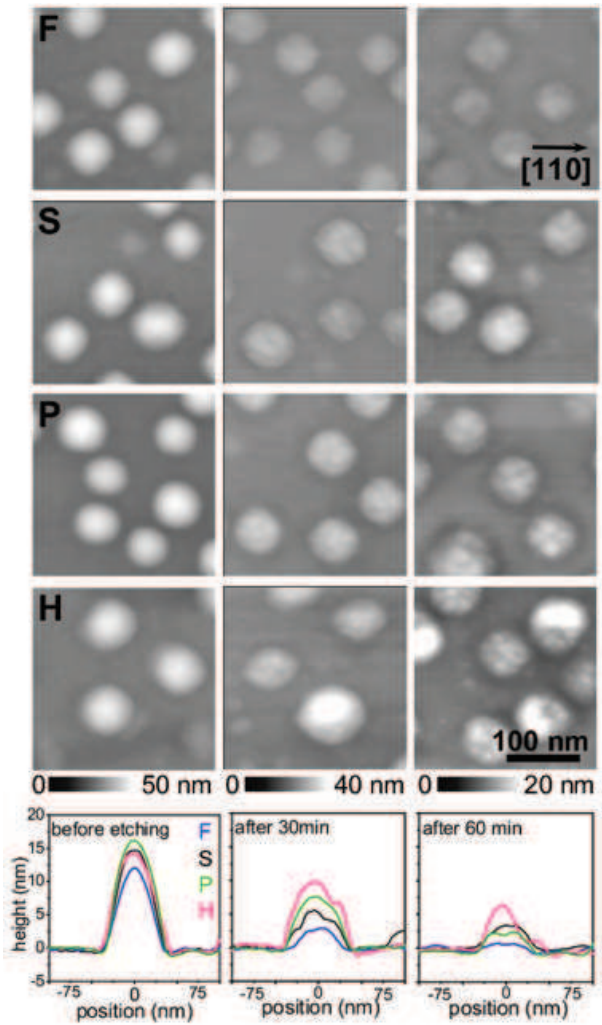}} \caption{AFM images and line
scans (bottom) of representative dome islands of all samples as
grown, after 30 min and after 60 min of etching with
25\%NH$_{4}$OH:31\%H$_{2}$O$_{2}$. All images were taken along
[110] direction.} \label{emap}
\end{figure}

Summarizing the above results, height statistics for samples as
grown and after etching for 30 and 60 min are shown in Figure 2. For
samples F, S, H and P the island heights are 12.4 $\pm$ 2.1 nm 14.3
$\pm$ 2.3 nm, 14.1 $\pm$ 2.8 nm, and 16.1 $\pm$ 2.3 nm,
respectively. The error bars correspond to the standard deviation
$\sigma$ of the island height distributions. The line scans (bottom
of Figure 1), together with the height statistics, show that for the
top 7-8 nm samples annealed P and H are richer in Si when compared
to samples as-deposited S and F. This situation is different for the
bottom 5 nm, where samples annealed P and as-deposited F exhibit a
Ge-rich base. From the evolution of $\sigma$ with etching, one finds
that sample annealed H has a consistently broader dome height
distribution than the others, indicating a wider composition range
within the island ensemble.

\begin{figure}[ht]
\centerline{\epsffile{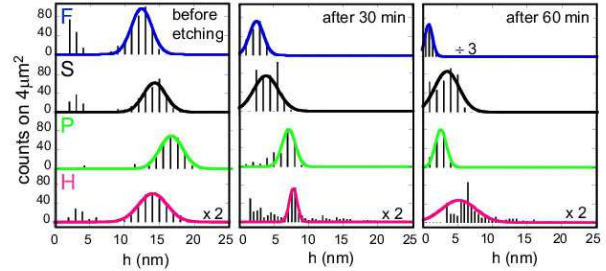}} \caption{Height statistics as
a function of etching time for all samples. } \label{emap}
\end{figure}

The sequence of line scans and height statistics shows that the
etching rate is not uniform, indicating that the Ge concentration in
the island is not constant in the growth direction. In Figure 3, the total integrated volume is shown as a
function of the etching time. The slope of the curve is associated
with the average Ge content in the film, with more material being
removed faster for Ge rich SiGe regions. Therefore, the Ge content
increases from sample H to S, to P and finally F. Comparing samples
as-deposited S and F shows that faster Ge deposition results in an
increase of the Ge content in the islands, corresponding to less
intermixing during growth\cite{Mcdaniel05}. While sample annealed H is richer in Si
compared to the reference sample (as-deposited S), annealing in
PH$_{3}$ enriches the Ge content compared to sample S. This
indicates that surface diffusion processes can be selectively
controlled depending on the proper choice of the annealing
environment.

\begin{figure}[ht]
\centerline{\epsffile{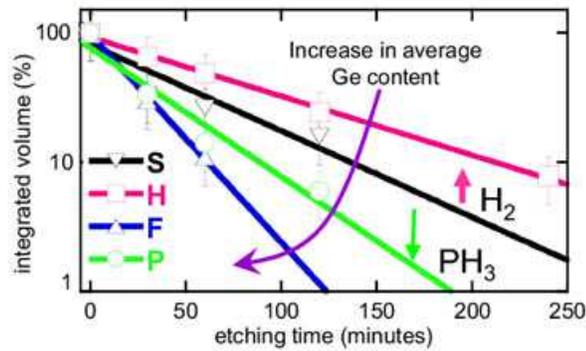}} \caption{Total integrated
volume (\%) as a function of the etching time (minutes) for all
samples. The slope of the curve is associated with the average Ge
content in the film, with
 material being removed faster for Ge rich SiGe regions.}
\label{emap}
\end{figure}

To understand the intra-island diffusion contribution to the alloy
formation, a more detailed analysis was carried out for sample
annealed P, where this particular process could be more clearly
isolated. The composition profile was studied by sequentially
etching and taking line scans along the [110] direction on the same
representative island. Figure 4(a) shows line scans for an island
before etching and after etching to h(x,y) $<$ 5nm. At this particular height, sample P exhibited a Ge content
higher than all samples except sample as-deposited F, indicating Ge
diffusion towards the island base (intra-island diffusion). The
region after etching shows a smaller diameter and height than prior
to etching, indicating a Ge rich shell. It also shows a small dip at
the center, corresponding to a Ge rich apex, similar to low
temperature MC simulations \cite{lang05}. In Figure 4(b), a 3D AFM
image exhibits a statistically significant rosette pattern, produced by atom redistribution within the island. This observation is associated with
strain-assisted intra-island diffusion. Similar morphology has been previously reported, also in
a post-growth annealing experiment\cite{Denker2003}, and was
attributed to enhanced surface diffusion, which occurred at the island edges. However, in the
present work, alloying takes place inside the island, and the edges remain
Ge rich, as can be seen in the line profiles of Figure 4(a); thus no Si could have come from the surface. The driving forces in
both experiments are basically the same - minimization of elastic
energy and maximization of entropy\cite{lang05,Ribeiro_2006}.

The Si-rich regions occur along the [110] directions. Since the
facet angle of the bounding \{311\} facets is less steep than the
\{15 3 23\} facets (25$^\circ$ and 32$^\circ$, respectively) a
smaller strain relaxation can take place in the [110] direction.
Therefore, Ge moves towards the soft [010] directions and Si moves
to the [110] direction, thus producing the rosette structure. At the
center of the rosette, one also finds a Ge rich region, presumably
reflecting more efficient relaxation at that site. The enhanced
intra-island diffusion occurs only very close to the substrate
(about 4 nm from the surface of an otherwise pristine island 16 nm
high), where the strain is large\cite{Paniago02}.

\begin{figure}[ht]
\centerline{\epsffile{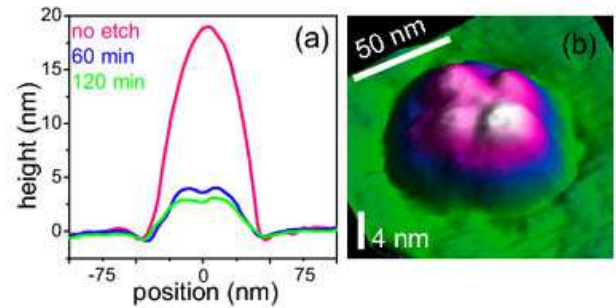}} \caption{ Sample annealed P:
{\bf(a)} Line scans on the same island as grown and after two
etching steps. {\bf(b)} 3D AFM image of one representative dome
after 60 minutes of etching, showing the rosette final
morphology.} \label{emap}
\end{figure}

In summary, a systematic study focused on the intermixing mechanisms
in Ge:Si(001) islands was carried out. Samples were grown and
annealed in different environments allowing different diffusion
processes to dominate. Selective etching and the RSM experiments
permitted a semi-quantitative picture of the Ge content profile
inside the dome islands. Increasing the growth rate decreased the
degree of Si-Ge intermixing. Intra-island diffusion occurred during
annealing in different environments, but surface diffusion could be
varied selectively by controlling the ambient gas. When Si surface
diffusion was minimized, atomic rearrangement took place within the
islands via intra-island diffusion, leading to a four-fold symmetric
rosette structure.

The authors would like to knowledge A. Rastelli and G. Katsaros for
fruitful discussion, and N. J. Quitoriano for his invaluable help in
RSM. The authors MSL and GMR thank FAPESP contract No. 03/09374-9,
CNPq and HPBrazil for financial support.

\end{document}